\documentclass[12pt,a4paper,twoside]{amsart}
\usepackage{txfonts}

\usepackage[author-year]{amsrefs}

\usepackage{amsthm}

\usepackage{graphicx}

\usepackage{algorithm}

\usepackage{algorithmic}
\usepackage{amsmath}

\theoremstyle{plain}

\theoremstyle{definition}

\theoremstyle{remark}

\makeatletter

\def\ps@plain{\ps@empty
  \def\@oddfoot{\normalfont\scriptsize \hfil\hfil}%
  \let\@evenfoot\@oddfoot}
\def\ps@headings{\ps@empty
  \def\@evenhead{%
    \setTrue{runhead}%
    \normalfont\large
    \rlap{}\hfil \leftmark{}{}\hfil}%
  \def\@oddhead{%
    \setTrue{runhead}%
    \normalfont\large \hfil
    \rightmark{}{}\hfil \llap{}}%
  \let\@mkboth\markboth
}
\def\ps@firstpage{\ps@plain
  \def\@oddfoot{\normalfont\scriptsize \hfil\hfil
     \global\topskip\normaltopskip}%
  \let\@evenfoot\@oddfoot
  \def\@oddhead{\@serieslogo\hss}%
  \let\@evenhead\@oddhead 
}

\def\@setaddresses{\par
  \nobreak \begingroup
\large
  \def\author##1{\nobreak\addvspace\bigskipamount}%
  \def\\{\unskip, \ignorespaces}%
  \interlinepenalty\@M
  \def\address##1##2{\begingroup
    \par\addvspace\bigskipamount\indent
    \@ifnotempty{##1}{(\ignorespaces##1\unskip) }%
    {\scshape\ignorespaces##2}\par\endgroup}%
  \def\curraddr##1##2{\begingroup
    \@ifnotempty{##2}{\nobreak\indent{\itshape Current address}%
      \@ifnotempty{##1}{, \ignorespaces##1\unskip}\/:\space
      ##2\par}\endgroup}%
  \def\email##1##2{\begingroup
    \@ifnotempty{##2}{\nobreak\indent{\itshape E-mail address}%
      \@ifnotempty{##1}{, \ignorespaces##1\unskip}\/:\space
      \ttfamily##2\par}\endgroup}%
  \def\urladdr##1##2{\begingroup
    \@ifnotempty{##2}{\nobreak\indent{\itshape URL}%
      \@ifnotempty{##1}{, \ignorespaces##1\unskip}\/:\space
      \ttfamily##2\par}\endgroup}%
  \addresses
  \endgroup
}

\def\@maketitle{%
  \let\@makefnmark\relax  \let\@thefnmark\relax
  \ifx\@empty\@date\else \@footnotetext{\@setdate}\fi
  \ifx\@empty\@subjclass\else \@footnotetext{\@setsubjclass}\fi
  \ifx\@empty\@keywords\else \@footnotetext{\@setkeywords}\fi
  \ifx\@empty\thankses\else \@footnotetext{%
    \def\par{\let\par\@par}\@setthanks}\fi
  \@mkboth{\@nx\shortauthors}{\@nx\shorttitle}%
  \global\topskip42\p@\relax 
  \@settitle
  \ifx\@empty\authors \else \@setauthors \fi
  \ifx\@empty\@dedicatory
  \else
    \baselineskip18\p@
    \vtop{\centering{\large\itshape\@dedicatory\@@par}%
      \global\dimen@i\prevdepth}\prevdepth\dimen@i
  \fi
  \@setabstract
 \normalsize
  \if@titlepage
    \newpage
  \else
    \dimen@34\p@ \advance\dimen@-\baselineskip
    \vskip\dimen@\relax
  \fi
}

\long\def\@footnotetext#1{%
  \insert\footins{%
\normalsize
    \interlinepenalty\interfootnotelinepenalty
    \splittopskip\footnotesep \splitmaxdepth \dp\strutbox
    \floatingpenalty\@MM \hsize\columnwidth
    \@parboxrestore \parindent\normalparindent \sloppy
    \protected@edef\@currentlabel{%
      \csname p@footnote\endcsname\@thefnmark}%
    \@makefntext{%
      \rule\z@\footnotesep\ignorespaces#1\unskip\strut\par}}}

  \def\@oddhead{\thepage\hfil\textsc{ Kateryna Mishchenko, Volodymyr
Mishchenko and Anatoliy Malyarenko}\hfil}
  \def\@evenhead{\hfil\textsc{Adapted Downhill Simplex
 Method\dots}\hfil\thepage}

\renewenvironment{abstract}{%
  \ifx\maketitle\relax
    \ClassWarning{\@classname}{Abstract should precede
      \protect\maketitle\space in AMS documentclasses; reported}%
  \fi
  \global\setbox\abstractbox=\vtop \bgroup
    \normalfont\normalsize
    \list{}{\labelwidth\z@
      \leftmargin3pc \rightmargin\leftmargin
      \listparindent\normalparindent \itemindent\z@
      \parsep\z@ \@plus\p@
      
    }%
    \item[\hskip\labelsep\scshape\normalsize\abstractname.]%
}{%
  \endlist\egroup
  \ifx\@setabstract\relax \@setabstracta \fi
}

\def\@captionfont{\large\normalfont}

\makeatother

\begin{document} {\large

{\small\hfill Research Reports MdH/IMa}

{\small\hfill No. 2007-5, ISSN 1404-4978}

\vspace{15mm} \addtolength{\topmargin}{-0.7cm}

\title{Adapted Downhill Simplex
 Method for Pricing Convertible Bonds}

\author[]{Kateryna Mishchenko}
\address{Department of Mathematics and
Physics, M\"{a}lardalen University, Box 883, SE-721 23
V\"{a}ster{\aa}s, Sweden}
 \email{kateryna.mishchenko@mdh.se}

\author[]{Volodymyr Mishchenko}
\address{Master Student at the Department of Numerical Analysis and Computer Science, Royal Institute of Technology, Stockholm, Sweden}
 \email{vladimir\_mishchenko@yahoo.com}
\author[]{Anatoliy Malyarenko}
\address{Department of Mathematics and
Physics, M\"{a}lardalen University, Box 883, SE-721 23
V\"{a}ster{\aa}s, Sweden}
 \email{anatoliy.malyarenko@mdh.se}
\date{\today}
\thanks{{The authors are grateful to Dr. Axel Kind,
Swiss Institute of Banking and Finance University of St. Gallen, for
his comments on the economical aspects of the study, in particular,
on the nature of convertible bonds. This was greatly significant
while writing this paper.}}

\begin{abstract}
The paper is devoted to modeling optimal exercise strategies of the
behavior of investors and issuers working with convertible bonds.
This implies solution of the problems of stock price modeling,
payoff computation and min-max optimization.

Stock prices (underlying asset) were modeled under the assumption of
the geometric Brownian motion of their values. The Monte Carlo
method was used for calculating the real payoff which is the
objective function. The min-max optimization problem was solved
using the derivative-free Downhill Simplex method.

The performed numerical experiments allowed to formulate
recommendations for the choice of appropriate size of the initial
simplex in the Downhill Simplex Method, the number of generated
trajectories of underlying asset, the size of the problem and
initial trajectories of the behavior of investors and issuers.

\end{abstract}
\maketitle \section{Introduction and Problem Formulation}
\subsection{Convertible Bonds}

\ One type of securities at modern financial market is a convertible
bond. They belong to most popular securities of modern financial
market. This type of securities is of interest for both small
developing companies for attracting investments and for investors,
 since for the latter such bonds are highly profitable.

Strategies of the behavior of investors and issuers for early
exercise decision must be chosen in such a way that the issuers'
payoff will be minimal while the investors' profit will be maximal.

A standard convertible bond is a bond that gives the holder
(investor) the right to exchange (convert) it into a predetermined
number of stock during a certain, predetermined period of time
\cite{ref1}.

\ Convertible bonds are characterized by the following options:

\emph{\textbf{ The Issuers Options}}
\begin{itemize}
\item[\textbf{1.}] \textbf{Call price $K_{t}$}

\ This option allows the issuer to call back the convertible bond at
the time $t$ with the payment $K_{t}$ to the investor.

\item[\textbf{2.}] \textbf{Call Notice Period $time_{notice}$}

\ Before calling back the convertible bond the issuer announces his
intend and can call convertible bond only after the call notice
period $time_{notice}$. During this period the investor may convert
the bond ("force conversion").
\end{itemize}

\emph{\textbf{The Investors Options}}
\begin{itemize}
\item[\textbf{1.}] \textbf{Number of stocks $n$}

\ The investor may convert the bond into $n$ stocks at any time
during the predetermined period.
\item[\textbf{2.}] \textbf{Put price $P_{t}$}

\ This option allows the investor to sell the convertible bond at
the price $P_{t}$ during the predetermined period.
\end{itemize}

\emph{\textbf{Common Options of Convertible Bonds and Stocks}}
\begin{itemize}
\item[\textbf{1.}] \textbf{Maturity Time $T$}

\ The expiry time of a convertible bond.
\item[\textbf{2.}] \textbf{Face Value of Convertible Bond $N$}

\ The predetermined price of a convertible bond which the issuer
will pay to the investor at maturity time.
\item[\textbf{3.}] \textbf{Redemption Ratio $\kappa$}

\ This is a preset percentage of the face value of a convertible
bond which increases the price of the face value. So, the $\kappa N$
instead of $N$ may be paid by the issuer to the investor. Usually
$\kappa$ is equal to 1.
\item[\textbf{4.}] \textbf{Price of Coupon Bond $B_{t}$}

\ This is the premium paid by the issuer to the investor at some
fixed time moments during the preset period.
\item[\textbf{5.}] \textbf{Value of continuation $V_{t}$}

\ This is the price of a convertible bond at every time moment
during the period when this convertible bond is alive. $V_{t}$ is
valued by the amount of money which the owner may get by converting
the bond into stocks.
\item[\textbf{6.}] \textbf{Current Stock price $S_{t}$}

\ This is the current price of stocks owned by the issuer.
\item[\textbf{7.}] \textbf{One Option}

\ The issuer or investor can perform only one action with the
convertible bond, i.e. if the investor converts the bond the latter
expires, as well as if the issuer wants to call the bond this cannot
be stopped.
\end{itemize}

\ Finally, the initial conditions at the time $t = 0$ are:
\begin{itemize}
\item[\textbf{1.}] $K_{t} > N$

\ The call price of a convertible bond is greater than its face
value.
\item[\textbf{2.}] $P_{t} < N$

\ The put price of a convertible is less than its face value.
\item[\textbf{3.}] $n = N/(S_{0}\cdot \eta)$, where $\eta > 1$.

\ $n$ is the number of stocks obtained by conversion of a bond. This
value is calculated under assumption that the stock price at the
initial moment is greater than the real one.
\end{itemize}
\subsection{Problem Formulation}\label{sec3}
\par Objective function is as a payoff gained by the
investor. This means that the payoff is a nonnegative number. \ It
is clear that the investor wants to maximize the payoff, while the
the issuer wants to minimize it. This payoff is based on the
behaviors of the investor and the issuer and on current stock price
at the time $t$.
\\ The
problem under consideration is to find such strategies for the
investor $Conv_{t}$ and the issuer  $ Call_{t}$ which maximize the
investor's payoff and minimize the payoff payed by the issuer to
investor, simultaneously. In other words, we have a min-max
optimization problem for computing the issuer's and investor's
strategies.
\begin{equation}
\max_{Conv_{t}} \min_{Call_{t}}
\textbf{payoff}(Conv_{t},Call_{t})\label{eq0}
\end{equation}
 The natural choice of the boundary
conditions for the investor's and issuer's strategies is the
following:
\begin{eqnarray}
Conv_{t} \geq N \label{eq1}\\
K_{t}\geq Call_{t}\geq N \label{eq2}
\end{eqnarray}

\ We consider (\ref{eq2}) only under the condition $T - t >
time_{notice}$ because otherwise the investor will not have enough
time to realize his right to call back a convertible bond.
\\
\ Additionally, we introduce some initial settings, where we assume
that we have a zero coupon convertible bond (no coupon payment
during the preset period is done: $B_{t} = 0$).

\par Also, we use the following preset parameters:

\fbox{%
\parbox[t]{10cm}{%
\begin{algorithmic}\label{box1}

\STATE $N = constant$  

\STATE

\STATE $\kappa$ = 1 

 \STATE

\STATE $(K_{t} = constant)$ and $(K_{t} > N)$  

\STATE

\STATE $S_{0} = constant$. ($S_{0}\cdot n < N$)  

\STATE

\STATE $\eta > 1$, say $\eta = 1.1$

\STATE

\STATE $n = N/(S_{0}\cdot \eta)$ 

\STATE

\STATE Initial Guess: $Conv_{t} = K_{t} + \varepsilon$; $Call_{t} =
N + \varepsilon$

\STATE

\STATE $time_{notice} = const$  

\end{algorithmic}

} }

\

 The maximization problem (\ref{eq0}) - (\ref{eq3}) will be solved by the global
 Downhill Simplex method, see \cite{ref3} and stock prices are
 modeled by Monte Carlo simulations presented in \cite{ref2}.

 The fact that such a problem can be solved numerically using this approach was shown
in the Master Thesis \cite{ref25}.

 Section 2 of this paper is devoted to the description of the methods for stock
 price generating, payoff computation, approximations of the
 strategy of the behaviors of the investor and the issuer, as well
 as the description of the Downhill Simplex method.

In Section 3 we present the results of the numerical experiments
with model under consideration. In these experiments we try to
determine the best values for such input parameters as size of the
simplex in Downhill Simplex method, the best choice for the initial
trajectories, the optimal of trajectories used for the stock price
generation and the appropriate number of the points for trajectories
approximations.

We finalize our work by making conclusions and giving some
guidelines for further investigation in Section 4.

\section{Numerical Issues}

\subsection{Stock Price Generating}\label{sec21}

\ The first step in solving the problem (\ref{eq0})-(\ref{eq2}) is
to generate stock prices. This can be done by different methods, and
we base our computation on the method producing the Brownian motion
type trajectories (see e.g. \cite{ref2}). This method generates
trajectories without jumps.
\\The initial stock price $S_{0}$ is given. The formula for
generating the stock price without jumps at $t+1$ time moment is:
\begin{equation}
S_{t+1} = S_{t}\cdot e^{(r - \delta - \sigma^{2}/2)\cdot \triangle +
\sigma\cdot\sqrt{\triangle}\cdot Z_{t + 1}} \label{eq3}
\end{equation}
 \ where $S_{t}$ is the
stock price at the current time moment $t$;
 $r$ is the
interest rate; $\delta$ is the dividend yield of the issuer stocks
(underlying asset); $\sigma$ is the volatility; $\triangle =
\frac{1}{250}$ is the time step at 250 working days a year and
$\{Z_{t+1}, t\ge 0\}$ is a sequence of independent standard normal
random variables.

\subsection{Computation of the Payoff}

\ We use the Monte Carlo method for modeling a real payoff. For this
purpose we generate M, a large number of trajectories, of the issuer
stock prices according to (\ref{eq3}), and then compute $payoff_{i}$
for each of them according to Algorithm \ref{alg1}.

\begin{algorithm}[htbp]
\caption{Payoff (objective function) Computation}\label{alg1}
\begin{algorithmic}
\STATE

\STATE $payoff = 0$

\STATE $flag_{notice} = 0$

\STATE $time_{check} = 0$

\STATE

\FOR {$t = 0:T$ $(T = maturity\cdot 250)$}

\STATE

\IF{$S_{t} > Conv_{t}$}
    \STATE $payoff = S_{t}$  (conversion or force conversion during the notice period)
    \STATE $break$
\ENDIF

\IF{$flag_{notice} = 1$}
    \IF{$time_{check} = time_{notice}$}
        \STATE $payoff = K_{t}$   (call)
        \STATE $break$
    \ELSE
        \STATE $time_{check} = time_{check} + 1$
    \ENDIF
\ELSE
    \IF{$(S_{t} > Call_{t})$ and $((T - t) > time_{notice})$}
        \STATE $flag_{notice} = 1$    (start call notice period)
    \ENDIF

\ENDIF

\IF{$t = T$}
    \STATE $payoff = N$ (the face value)
\ENDIF

\ENDFOR

\end{algorithmic}
\end{algorithm}

\ The put price $P_{t}$ is less than the face value $N$. In the case
of maximization of the investor's payoff we will not take into
account $P_{t}$ and suppose that the investor does not use the
possibility of choosing the put option.

\ Finally, the real payoff is:
\begin{equation}
payoff = \frac{1}{M}\cdot\sum_{i = 1}^{M} payoff_{i} \label{eq4}
\end{equation}

\ In this study we simplify the problem by setting $r = const$;
$\delta = const$ and $\sigma = const$.

\subsection{Optimization procedure by Downhill Simplex Method}

\ As we solve a min-max optimization problem (\ref{eq0}), the
optimization procedure is to be applied twice at each iteration:
once as a maximizer of the objective w.r.t. the investor's strategy
and then as a minimizer w.r.t. the issuer's strategy.

\ In other words, we find a pair of trajectories $ (Conv_{t}^{*},
{Call_{t}}_{*})$ which satisfy the conditions:
\begin{eqnarray}
 \text{payoff}(Conv_{t}^{*},Call_{t})= \max_{Conv_{t}}
\textbf{payoff}(Conv_{t},Call_{t}) \label{eq10}\\
\text{payoff}(Conv_{t}^{*},{Call_{t}}_{*}) = \min_{Call_{t}}
\textbf{payoff}(Conv_{t}^{*},Call_{t})\label{eq11}
\end{eqnarray}

\ The optimization problem under consideration is very computational
expensive due to the Monte Carlo simulations used of evaluation of
the objective function, thus we do not apply the optimization
methods based on derivatives computation.

 To perform the optimization procedure in the most efficient way we use a derivative-free Downhill Simplex method \cite{ref3},
 which is also easy implementable.

 Below we give a description of this algorithm for minimizing some function $f(x)$.

\ Firstly we need to specify the initial $m+1$ dimensional simplex
by taking $m$ points around the initial guess $x^1$. The point with
the highest function value $f_{max}$ is called $x_{max}$.

\ The main idea of the Downhill Simplex method is to substitute
 a point with the coordinates $x_{max}$ by another point with
better function value (e.g. get lower $f_{max}$ in case of
minimization problem). This is done by means of Reflection,
Expansion, Contraction and Multiple Contraction. Let consider these
functions shortly.

\ Firstly, we introduce the point $x_{center}$  (the center point of
the simplex for current iteration):
\begin{equation}
x_{center} = \frac{1}{m+1}\sum_{i = 1}^{m+1}{x^{i}}\label{eq5}
\end{equation}

\begin{itemize}
\item[\textbf{1.}] \textbf{Reflection}

\ The point $x_{max}$ is reflected into $x_{max}^{*}$ so that it
lies on the opposite (to the point $x_{max}$) side of the line
containing $x_{max}$ and $x_{center}$. The distance between
$x_{center}$ and $x_{max}^{*}$ depends on a positive constant
$\alpha$ - \emph{reflection coefficient} and is computed as
\begin{equation}x_{max}^{*} = (1 +
\alpha)\cdot x_{center} - \alpha\cdot x_{max}\label{eq6}
\end{equation}

\item[\textbf{2.}] \textbf{Expansion}

\ The expansion process is a prolongation of Reflection, and the new
point $x_{max}^{**}$ is found as:
\begin{equation}
x_{max}^{**} = \gamma\cdot x_{max}^{*} + (1-\gamma)\cdot
x_{center}\label{eq7}
\end{equation}
where $\gamma  (\gamma > 1)$ is the \emph{expansion coefficient}.

\item[\textbf{3.}] \textbf{Contraction}

\ The contraction process puts the next point between $x_{max}$ and
$x_{center}$, and the new point $x_{max}^{*}$ can be found as:
\begin{equation}
x_{max}^{*} = \beta\cdot x_{max} + (1-\beta)\cdot
x_{center}\label{eq8}
\end{equation}
where $0<\beta<1$ is the \emph{contraction coefficient};

\item[\textbf{4.}] \textbf{Multiple Contraction}

\ Here all the points are shifted according to the following rule:
\begin{equation}
x^{i} = (x^{i} + x_{min})/2, \ i = 1,..m+1 \label{eq9}
\end{equation}
where $i$ is the number of points in the simplex and $x_{min}$
corresponds to the point with the minimal function value $f_{min}$.
\end{itemize}

In Algorithm \ref{alg2} we implement the Downhill Simplex method
described in \cite{ref4}.

\begin{algorithm}[htbp]
\caption{Downhill Simplex Method}\label{alg2}
\begin{algorithmic}

\STATE Choose the initial guess $x^{1}$

\STATE Choose the size of initial simplex $k$

\STATE Choose the maximum number of iterations $max_{iter}$

\STATE Choose the termination criterion $\epsilon$

\STATE

\FOR {$i = 2:m+1$}

   \STATE $x^{i} = x^{1}$

   \STATE $x_{i}^{i} = x_{i-1}^{i} + k$

   \STATE $f(i) \equiv f(x^{i})$

\ENDFOR

\STATE MAIN LOOP

\FOR {$i = 1:max_{iter}$}

   \STATE Find $f_{max}$, $f_{min}$, and $f_{nearest_{max}}$

   \STATE

   \IF {(($i == max_{iter}$) or ($(f_{max} - f{min}) < \epsilon$))}

      \STATE $break$

   \ENDIF

   \STATE

   \STATE \textbf{Reflection} $x_{max}$ to $x_{max}^{*}$, and find $f_{max}^{*}$

   \STATE

   \IF {$f_{max}^{*} < f_{min}$}

      \STATE \textbf{Expansion} $x_{max}^{*}$ to $x_{max}^{**}$ and find the new function value  $f_{max}^{**}$

      \STATE

      \IF{$f_{max}^{**} < f{max}^{*}$}

         \STATE $f_{max}^{*} = f_{max}^{**}$, $x_{max}^{*} = x_{max}^{**}$

      \ENDIF

   \ELSE

      \IF{$f_{max}^{*} > f_{nearest_{max}}$}

         \STATE \textbf{Contraction} $x_{max}$ to $x_{max}^{*}$, and find $f_{max}^{*}$

         \STATE

      \ENDIF

   \ENDIF

   \STATE

   \IF {$f_{max}^{*} > f_{max}$}

      \STATE \textbf{Multiple Contraction}

   \ELSE

      \STATE $x_{max} = x_{max}^{*}$, $f_{max} = f_{max}^{*}$

   \ENDIF

\ENDFOR

\end{algorithmic}
\end{algorithm}

We denote as $x^i$ the whole strategy on $i$th iteration consisting
of all $j$ points and  $x_j$ is the $j$ component of the strategy
$x^i$.

\ So, for solving the maximization problem (\ref{eq10}) we use the
Algorithm \ref{alg2}, where we minimize the objective function $-f$.
The problem (\ref{eq11}) is solved by Algorithm \ref{alg2} directly.
\subsection{Dimension of the Problem and Investor(Issuer) Trajectory
Approximation}\label{subsec23}

\ The trajectories $Conv_{t}$ and $Call_{t}$ are the vectors of
length $250\cdot Maturity$ and $250\cdot Maturity -
CallNoticePeriod$, respectively. Since the lengthes of these vectors
are the dimension of the optimization problem, it is clear that such
a problem cannot be solved efficiently by any optimization method.
In order to reduce the dimension of the problem, solved by the
Downhill Simplex Method we shall consider some approximations of the
trajectories instead of the original extremely costly computable
trajectories.

So, instead of considering whole trajectories $Conv_{t}$ and
$Call_{t}$ in optimization procedure described above, we use a set
of threshold points $x_i$- the critical dates from the first
possible exercise date till last exercise date at maturity $T$. The
trajectories are approximated by piecewise linear functions with
nodes being the threshold points $x_i$.
\par We will use $5-15$ points approximation of the trajectories
which means that the optimization problem will be $5-15$
dimensional, too.

\ Since we have a Brownian type modeling, the deviations of the
prices from the initial state will increase while approaching the
maturity. This is also natural for any market that the most
interesting and important actions take place at the end. So, the
distributions of the points should meet this requirement. According
to \cite{ref25} we consider the following distribution of the
threshold points:
\begin{equation}
x_{i+1} = \frac{x_{m}- x_{i}}{2}+ x_{i}, \ i = 1,..m-2\label{eq13}
\end{equation}
where $x_{1}= 0$ and $x_{m} = T$ or $x_{m}= T -time_{notice}-1$ for
the investor's and issuer's trajectories accordingly. Also $m$ is
the number of points in approximation.
\\ Note that for the investor's trajectory the function value at the point $x_{m} = T$
is always equal to $N$.

\subsection{Short Description of the Main
Algorithm} As a termination criterion we use the value of the gap.
The gap is the difference between the investors and issuers payoff.
For each iteration it is defined as:
\begin{equation}
\text{gap} =
\text{payoff}(Conv_{t}^{*},Call_{t})-\text{payoff}(Conv_{t}^{*},{Call_{t}}_{*})
\label{eq14}
\end{equation}
where $ payoff(Conv_{t}^{*},{Call_{t}})$ and
$payoff(Conv_{t}^{*},{Call_{t}}_{*})$ are the optimizers for the
problems (\ref{eq10})- (\ref{eq11}) respectively.

From the optimization point of view, due to the definition
(\ref{eq14}), the gap is always nonnegative.

The main algorithm for solution the problem (\ref{eq0}) -
(\ref{eq3}) is  sketched in the Algorithm \ref{alg3}.

\begin{algorithm}[H]
\caption{Body of the main loop}\label{alg3}
\begin{algorithmic}

\STATE

\STATE $gap = gap_{old} = 10$

\WHILE{(($gap > \epsilon$) and ($gap_{old} > \epsilon$))}
  \STATE $gap_{old} = gap$

  \STATE compute $payoff(Conv_{t}^{*},{Call_{t}})$ from (\ref{eq10}) by one step of
   Algorithm \ref{alg2}

  \STATE compute $payoff(Conv_{t}^{*},{Call_{t}}_{*})$ from (\ref{eq11}) by one step of Algorithm
  \ref{alg2}

  \STATE   $gap = payoff(Conv_{t}^{*},{Call_{t}}) - payoff(Conv_{t}^{*},{Call_{t}}_{*})$
\ENDWHILE

\end{algorithmic}
\end{algorithm}

\ where $\epsilon$ is a small constant, say $\epsilon = 0.0001$.

\section{Numerical Results}

\par The numerical model described above has a quite complicate structure and
is sensitive to the choice of the input parameters. In this section
we investigate the dependence of the performance of the method on
some of parameters. This analysis is used to validate the numerical
model and choose the parameters which give a reasonable and fast
solution.
\par In each experiment we compute and compare the
optimal strategies of the investor and the issuer. Moreover, we
present the history of the behavior of the optimal value of the
objective, i.e. the investor's payoff.

\ While comparing the results of the numerical experiments with
different initial parameters it is necessary to assume that due to
the specificity of the given optimization problem the following
conditions are to be fulfilled:
\begin{itemize}
\item The strategies of the issuer and the investor must change their behavior near
the maturity time, see Section \ref{subsec23};
\item The objective function must produce dumping oscillations due to the nature of the $min-max$ problem.
\end{itemize}

\ Below we present the experiments where 4 parameters (initial
conditions) of the optimization problem are varied. These parameters
are: the number of generated trajectories, the size of the simplex,
the initial guess and the number of points for approximation
strategies of the investor and the issuer (the size of the problem).
\par In each experiment
the optimization problem is solved 3 times for 3 experimental values
for each of the four parameters, three other parameters being fixed
(are from the basic set).
\par The basic set of the parameters is the following:
\begin{eqnarray}
\text{number of generated trajectories M} & = & 525 \label{incon1} \\
 \text{size of simplex k}  & = & 3  \label{incon2}\\
\text{size of the problem m} & = & 10 \label{incon3}\\
 \text {initial guess}\  \varepsilon & = & 5 \label{incon4}
\end{eqnarray}
 \ The rest of the parameters of the problem are constant values, such as:
\begin{itemize}
\item [] Initial Stock Price = \$98;
\item [] Convertible Bond Price (Face value) = \$100;
\item [] Call Price = \$110;
\item [] Call Notice Period = 10 days;
\item [] Interest Rate = 0.05 (5\%);
\item [] Dividend yield = 0.1;
\item [] Volatility = 0.2 (20\%);
\item [] Maturity = 2 (two years).
\end{itemize}
\subsection{Experiment 1: Different number of trajectories for Stock
Price generating}\label{subsec31}
 \ Presented here are the results for 3
different sets of generated trajectories: $M = 50, 525$ and $1000$.
The rest of the parameters are (\ref{incon2}) - (\ref{incon4}).
%
%
%
%

\begin{figure}[H] \centering
\includegraphics[width=4.5in]{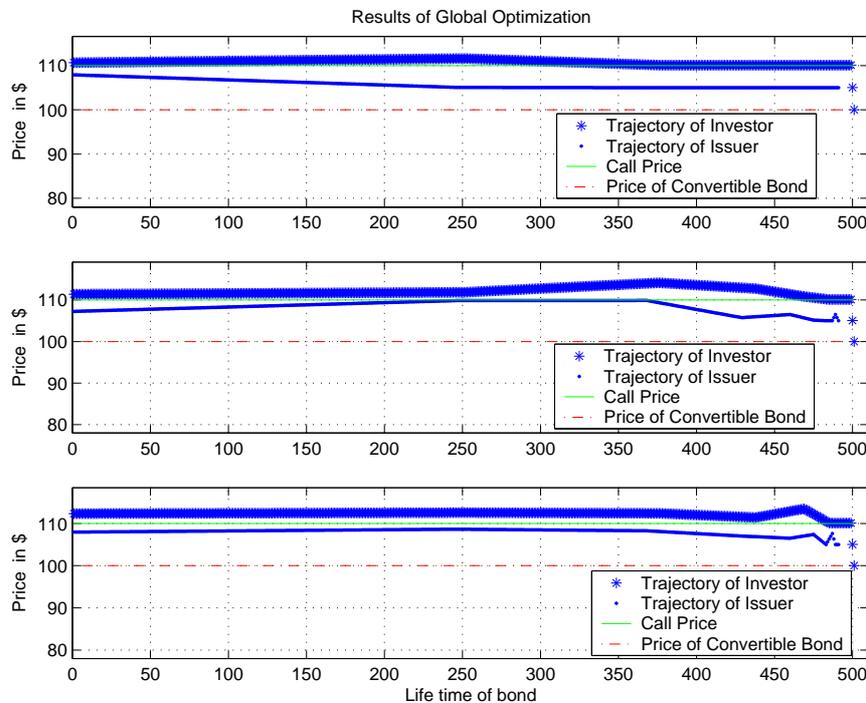}
\caption{Strategies of Investor and Issuer. Different number of
trajectories}\label{fig4}
\end{figure}

\begin{figure}[H] \centering
\includegraphics[width=4.5in]{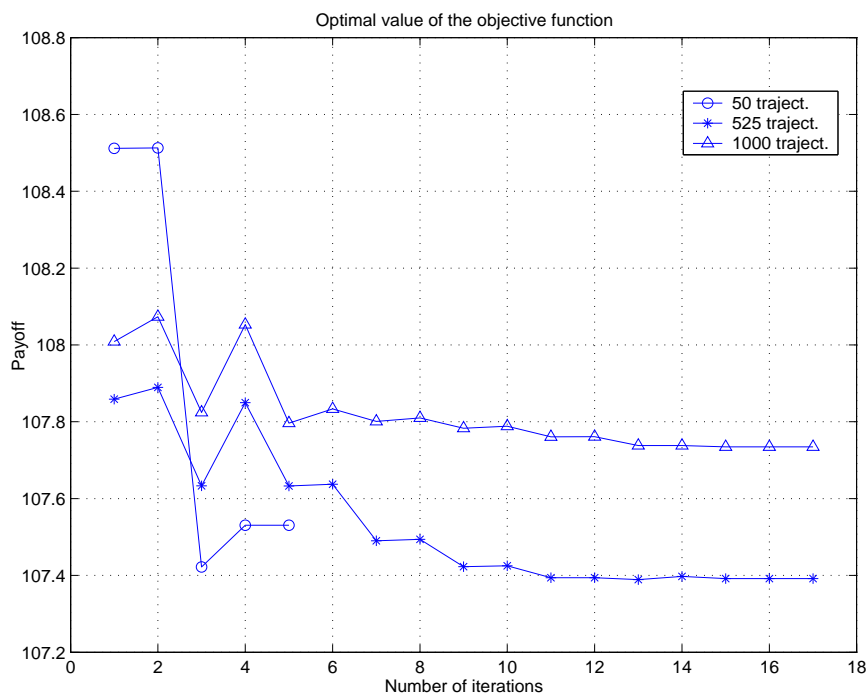}
\caption{Objective function history. Different number of
trajectories}\label{fig5}
\end{figure}

\par Analyzing the above figures, one can see that the solution corresponding to the case with 50
trajectories cannot be considered to be proper as the number of
trajectories is insufficient. Firstly, subplot 1 in Figure
\ref{fig4} shows that the behavior of the strategies close to the
maturity does not change. This means that the amount of generated
strategies has no real affect on the strategies of the investor and
the issuer at the end of the bond lifetime. Secondly, Figure
\ref{fig5} shows that the method terminates rather fast, which is
not appropriate for this min-max problem.
\par The behavior of the investor's and the issuer's optimal strategies as well as the
 objective function history are
similar in the cases with 525 and 1000 trajectories (see Figure
\ref{fig5} and subplots 2-3, Figure \ref{fig4}). Thus, these numbers
of generated stock price trajectories can be accepted for future
experiments. \ In the basic set (\ref{incon1}) - (\ref{incon4}) we
consider 525 trajectories since the time needed for function
evaluation for this case is much shorter than the one for the case
with 1000 trajectories, but the solution is acceptable thereat.

\subsection{Experiment 2: Different sizes of simplex}\label{subsec32}

\ This experiments concerns the proper choice of the parameter $k$,
which is the distance between the neighbor nodes in the initial
simplex used in Downhill Simplex method.
 This experiment we run for 3
sizes of the initial simplex: $k = 1, 3$ and $5$. The rest of the
parameters are (\ref{incon1}) and (\ref{incon3})- (\ref{incon4}).

\begin{figure}[htbp] \centering
\includegraphics[width=4.5in]{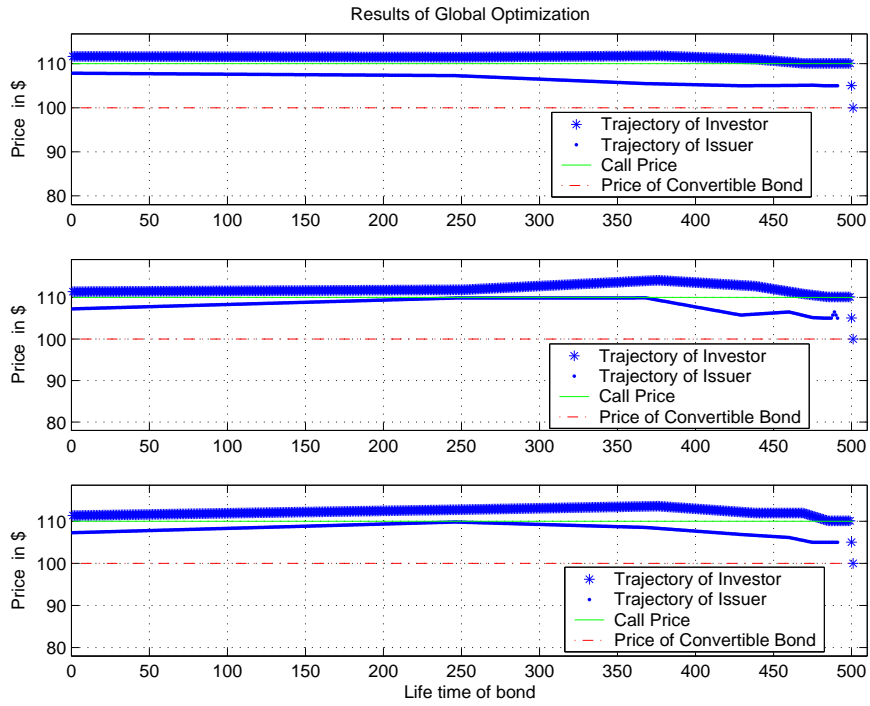}
\caption{Strategies of Investor and Issuer. Different sizes of
initial simplex}\label{fig7}
\end{figure}

\begin{figure}[htbp] \centering
\includegraphics[width=4.5in]{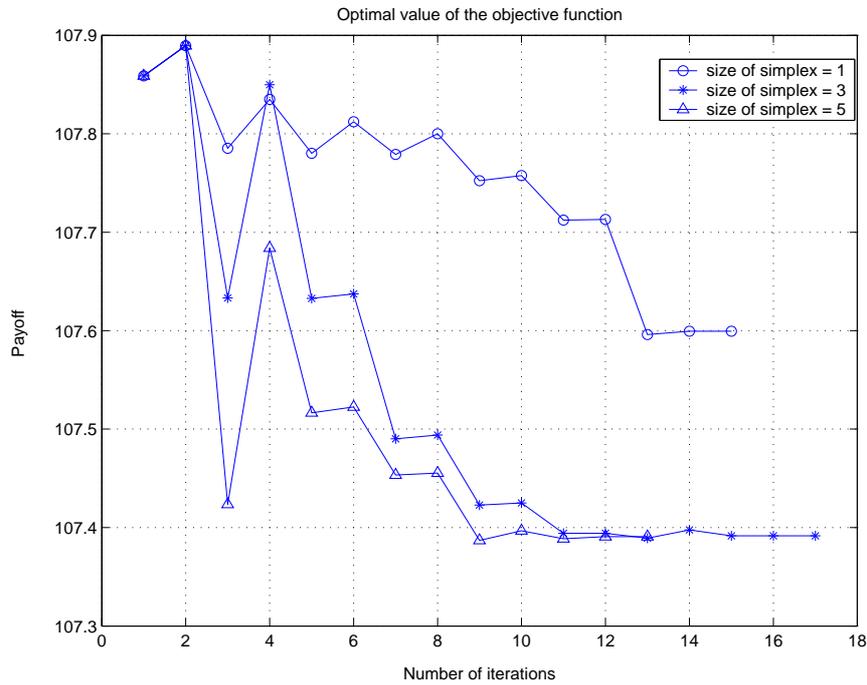}
\caption{Objective function history. Different sizes of initial
simplex}\label{fig8}
\end{figure}

\ Figure \ref{fig8} shows that for the minimal size of simplex $k =
1$ we have oscillation with small amplitude. So, the behavior of
strategies of the investor and the issuer do not change essentially
w.r.t the initial guess (see subplot 1, Figure \ref{fig7}).

\par On the contrary, for maximal size of simplex $k = 5$,
the first solution of the minimization has a dominating effect. In
other words, the first step down (the solution of the minimization
problem) has a big magnitude, which does not allow to produce
sufficiently big second step up (the solution of the maximization
problem). This effect manifests itself in subplot3, Figure
\ref{fig7}, where the functions corresponding to the strategies of
the investor and the issuer are flat near the maturity time.
\par So, we choose the size of simplex $k = 3$, which produces reasonable
steps for both strategies (see subplot 2, Figure \ref{fig7}).

\subsection{Experiment 3: Different initial trajectories}\label{subsec33}

\ For any global optimization procedure the initial guess, which is
the initial trajectories are of importance. We analyze 3 values of
the initial trajectories $\varepsilon = 1, 5$ and $9$. The rest of
the parameters are (\ref{incon1})-(\ref{incon3}).

\begin{figure}[htbp] \centering
\includegraphics[width=4.5in]{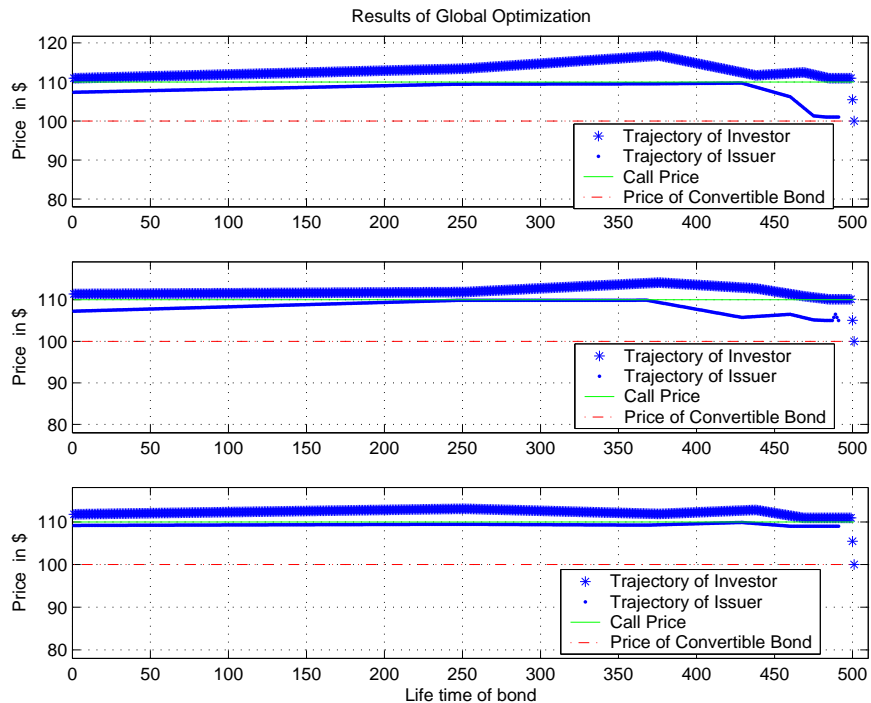}
\caption{Strategies of Investor and Issuer. Different initial
trajectories.}\label{fig10}
\end{figure}

\begin{figure}[htbp] \centering
\includegraphics[width=4.5in]{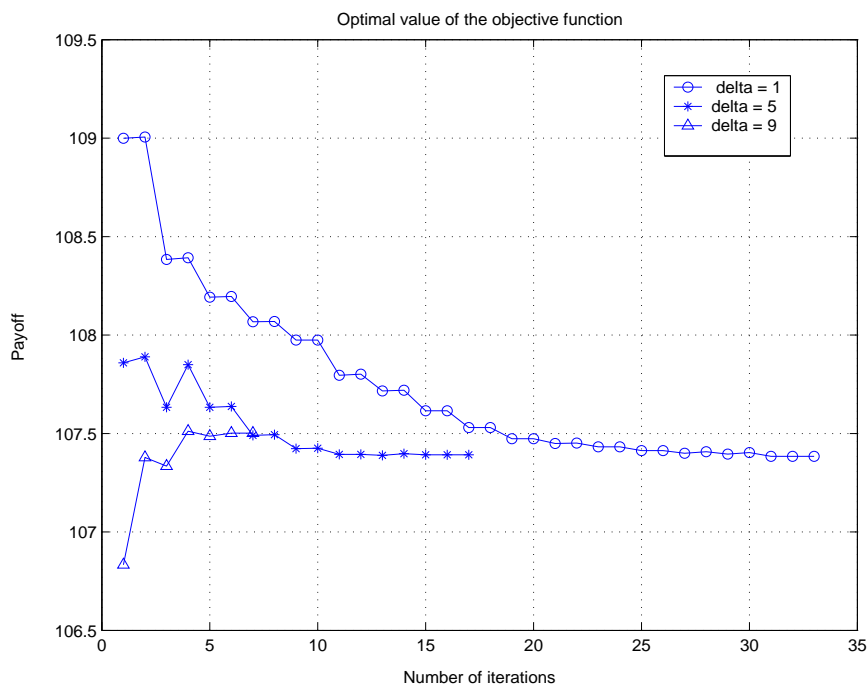}
\caption{Objective function history. Different initial
trajectories}\label{fig11}
\end{figure}

Figure \ref{fig11} shows that all three experiments terminate with
the same objective function value, but give different points
(strategies), see Figure \ref{fig10}.
\par The case with $\varepsilon = 1$
requires the maximal number of iterations (almost twice as many as
the case for $\varepsilon= 5$ and three times as many as for
$\varepsilon = 10$).
\par The case with $\varepsilon = 9$ seems to be not very informative,
since almost nothing happens close to the maturity (see Figure
\ref{fig10}, subplot 3).

\ So, the most interesting cases are $\varepsilon = 1$ and
$\varepsilon = 5$, but we choose $\varepsilon= 5$ in the basic set
because in this case the number of iterations is twice lower in
comparison with the case $\varepsilon = 1$, and it produces an
acceptable result.

\subsection{Experiment 4: Different sizes of the
problem}\label{subsec34}

\ Instead of using the whole trajectories for issuers and investors,
we used the approximated trajectories, see Section \ref{subsec23}.
The amount of points in (\ref{eq13}) which gives a reasonable
solution to the problem is the subject of investigation in this
experiment.

 We solved our problem for different sizes of the problem: $m = 5,
10$ and $15$. The rest of the parameters are
(\ref{incon1})-(\ref{incon2}) and (\ref{incon4}).

\begin{figure}[htbp] \centering
\includegraphics[width=4.5in]{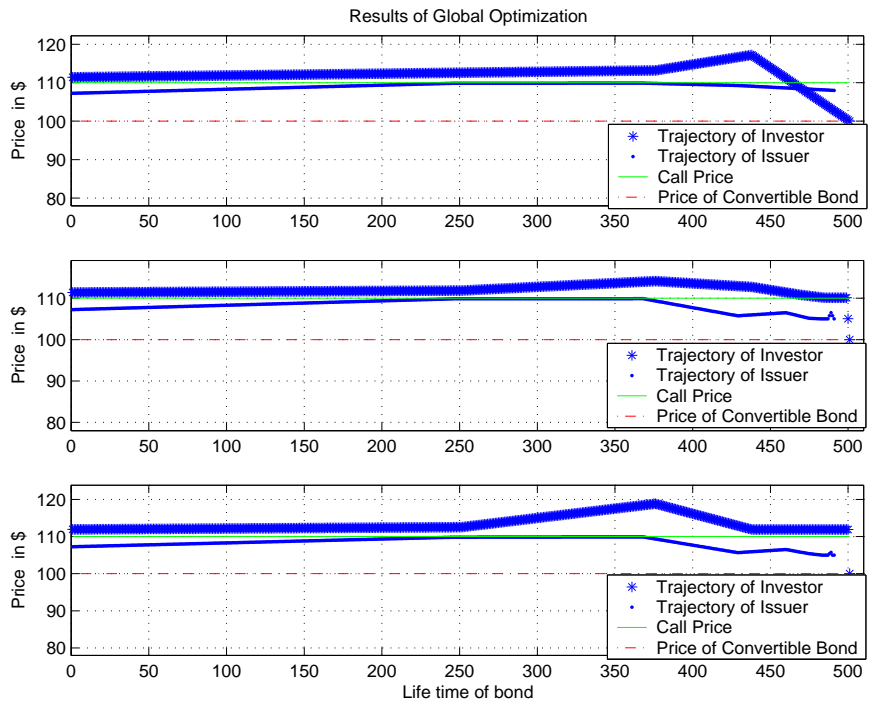}
\caption{Strategies of Investor and Issuer. Different problem
sizes}\label{fig13}
\end{figure}

\begin{figure}[htbp] \centering
\includegraphics[width=4.5in]{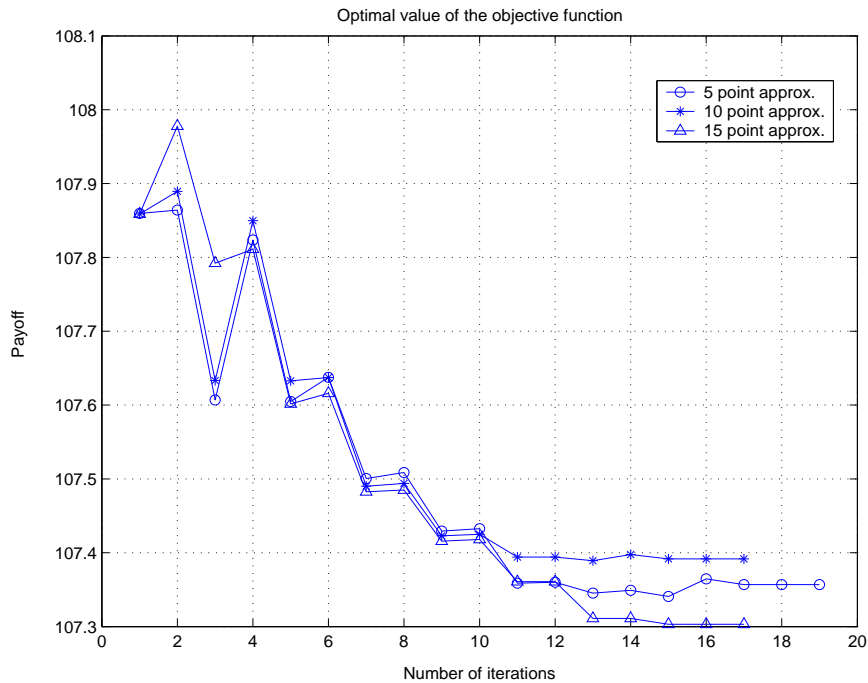}
\caption{Objective function history. Different problem
sizes}\label{fig14}
\end{figure}

\par Figure \ref{fig14} shows that the objective
function history for all the cases is almost similar.

Nevertheless, $5$-point approximation of the strategies is not
sufficient. Since the most interesting part of the strategy is the
second, it is not enough to have only 3 points for approximation of
the second part of the strategy. As seen from subplot 1, Figure
\ref{fig13}, the strategies are not smooth enough in the vicinity of
the maturity time.

\par $15$-point approximation gives very interesting results, but
the size of the problem becomes too high as well. So, the best
choice for the basic set is $10$-points approximation of the
strategies, which is the tradeoff between two other approximations.

\section{Conclusions and Suggestions for Further Investigation}

\ In this study we considered a method for computing the strategies
of the investor and the issuer dealing with convertible bonds. This
method consists of two main stages: stock price generating and
solution of the min-max optimization problem. For stock price
generating we used a Monte-Carlo method based on the formula
(\ref{eq3}), and applied the Downhill Simplex method for the
solution of the global nonlinear optimization problem. The results
of our investigation allow to draw the following conclusions.
\begin{itemize}
\item [1.] The proposed method is sensitive to the number of generated
trajectories of Stock Price. It means that for some small number of
generated trajectories the method does not produce any reasonable
solution. We suggest 500 trajectories as the minimal number required
for achieving a reasonable solution (see Section \ref{subsec31});
\item [2.] The Simplex Downhill method is sensitive to the size of the initial simplex.
It is very important to choose the initial simplex of a proper size,
otherwise there exists a risk to get non-acceptable solution, see
Section \ref{subsec32}. We recommend the size of simplex $k = 3$.
\item [3.] The Downhill Simplex is also sensitive to the choice of a good initial guess.
 The best choice in our experiments was  $\varepsilon = 5$ (see Section
\ref{subsec33});
\item [4.] The dimensions (size of the problem) is important
in our experiments, too. Very large size of the problem requires too
much computational time, but for a small size we get non-acceptable
solution. We took 10 points (solved $10$-dimensional problem), see
Section \ref{subsec34}.
\end{itemize}

\ For further investigation the following is to be taken into
account:

\par All the results presented in this study were obtained for
predetermined constant values such as initial stock price, call
price, face value of convertible bond, etc. So, it is very
interesting to run experiments with other values of the economic
parameters. Also, such parameters as volatility, interest rate,
dividend yield may vary during the lifetime of the bond. For
example, they may be recalculated every day, which will make stock
price generation more complicated. Finally, the problem may be
solved for nonzero coupon bond.
\par We used Brownian type stock price generation without jumps (see Section \ref{sec21}) which is one
of the options. It is possible to consider some other stock price
generation algorithms which have another nature (with jumps) and may
give an interesting effect on the results.
\par From the viewpoint of the optimization it would be extremely
useful to consider another global solver, since Downhill Simplex is
so sensitive to the choice of the initial point and the size of
initial simplex. On the other hand,  some local optimization methods
may be useful, since the problem is constrained and the feasibility
area is quit narrow.
\par For more efficient strategies approximation it may be very
helpful to consider other distribution of points (e.g. equidistant
distribution) and other ways of approximation, e.g. cubic splines.

}
\end{document}